\documentclass{article}
\usepackage[utf8]{inputenc}
\usepackage{braket,amsmath,graphicx,enumitem,bbm,hyperref,cleveref,amsthm,titlesec,cite,dsfont,subcaption,comment,appendix}
\newtheorem{example}{Example}[section]

\usepackage{booktabs}
\newcommand\headercell[1]{%
   \smash[b]{\begin{tabular}[t]{@{}c@{}} #1 \end{tabular}}}

\newcommand{\bbrakket}[2]{\langle\langle#1|#2\rangle\rangle}

\usepackage{xcolor}

\usepackage[nocomma]{optidef}

\title{Informationally Overcomplete POVMs for Quantum State Estimation and Binary Detection}
\author{Catherine Medlock$^1$, Alan Oppenheim$^1$, Petros Boufounos$^2$}
\date{%
    $^1$ MIT Department of Electrical Engineering and Computer Science\\%
    $^2$ Mitsubishi Electric Research Laboratories\\[2ex]%
    \today
}

\numberwithin{equation}{section}

\usepackage[refpage,intoc]{nomencl}
\usepackage{siunitx}
\usepackage{ifthen}

\makenomenclature

\usepackage{etoolbox}
\renewcommand\nomgroup[1]{%
  \item[\bfseries
  \ifstrequal{#1}{B}{Acronyms}{%
  \ifstrequal{#1}{A}{Symbols}{}}%
]}

\begin{document}

\maketitle


\section*{Preamble}

Our intention while preparing these notes has been for them to be readable and interesting to an audience with a wide range of backgrounds. We anticipate that some sections will be familiar to readers with a strong background in classical signal processing and particularly in frame theory, and other sections to readers with a strong background in quantum mechanics. It is our hope that both audiences will find the perspectives in the shared issues and overlap between the two fields to be interesting.

\section{Introduction} \label{sec:intro}

Quantum state detection -- the problem of identifying which of a given set of states accurately describes the state of an unknown system -- is an important problem in the fields of quantum information theory, quantum communication systems, and in the testing of quantum technologies. In a communication setting each possible state may represent a different transmitted message. When developing a device such as a quantum computer, the ability to read out the final state of a system after it has been processed is essential to evaluating the accuracy of the result. Theoretical results regarding detection strategies that minimize the probability of an error or maximize the mutual information between input and output, for example, are well-established \cite{helstrom1967detection,davies1978information}. Many of these results have also been verified experimentally (see \cite{chefles2000quantum} for references).

A typical formulation of the quantum state detection problem assumes that the final decision is made based on the outcomes of one or possibly multiple quantum measurements. As is well-known, a given quantum measurement can be mathematically modeled using a set of operators that collectively form a positive operator-valued measure (POVM). The elements of the POVM can be used to map the state of the system being measured to a sequence of probabilities, each of which corresponds to the probability of obtaining one of the possible measurement outcomes. An informationally complete (IC) POVM is one that maps each possible quantum state to a unique sequence of probabilities. An informationally overcomplete (IOC) POVM is, loosely speaking, an IC POVM for which the probability sequences contain some amount of redundancy. This redundancy can be beneficial in mitigating various sources of error that affect our estimations of the probabilities, as we describe further in Section \ref{sec:application_to_quantum}.

One of the most widely used ways of thinking about and analyzing IC POVMs relies on a fundamental result that establishes the equivalence of a given IC POVM with a complete (and possibly overcomplete) representation of an appropriate operator-valued vector space. This approach opens the door for the study of IC POVMs to leverage both the intuition and results from the field of frame theory, which can be broadly described as the study of overcomplete representations of finite- and infinite-dimensional vector spaces.

In these notes we start by reviewing the mathematical framework surrounding overcomplete representations of vector-valued vector spaces in Section \ref{sec:frame_representations}. The focus of Section \ref{ssec:robustness_of_frame_representations} is the problem of estimating an unknown vector in the presence of error on its frame coefficients. Under certain assumptions on the error and on the frame being used, it is well-known that there is a tradeoff between the total number of frame vectors and the magnitude of the individual error values, In Section \ref{sec:operator_spaces} we extend the disucssion of frame representations to include operator-valued vector spaces, or operator spaces for short. Operator spaces as they pertain to quantum mechanics are described in Section \ref{sec:operator_spaces_in_quantum_mechanics} and a fundamental result that connects IC POVMs to frames of a specific operator space is reviewed. In Section \ref{sec:application_to_quantum} we describe and demonstrate through simulations how under analogous assumptions to Section \ref{ssec:robustness_of_frame_representations}, the same tradeoff is present in the context of IC POVMs and quantum state estimation. Lastly we provide evidence through simulation that the tradeoff can also be exploited in the context of quantum binary state detection.

\section{Frame Representations} \label{sec:frame_representations}

The basic mathematical tools of frame theory are reviewed in Sections \ref{ssec:defn_of_a_frame} to \ref{ssec:dual_frames}. In Section \ref{ssec:robustness_of_frame_representations} we describe the robustness of frame representations to additive error on the frame coefficients of a given vector. The underlying motivation in presenting these topics is ultimately to apply them in the context of quantum mechancis. Thus, to be consistent with the quantum mechanics literature we use Dirac's bra-ket notation, in which a vector $x$ is represented by the ket $\ket{x}$ and its Hermitian conjugate is represented by the bra $\bra{x}$. The inner product between two vectors $\ket{x}$ and $\ket{y}$ is denoted as $\braket{x|y}$. For more details on bra-ket notation see, for example, Chapter 2 of \cite{nielsen2016quantum}.

\subsection{Definition of a Frame} \label{ssec:defn_of_a_frame}

We consider vectors that lie in an $N$-dimensional Hilbert space $\mathcal{V}$. Any set of vectors $\{\ket{f_k}, 1 \leq k \leq M\}$ that lie in and span $\mathcal{V}$ (and that may be linearly dependent) form what is referred to as a frame for $\mathcal{V}$. More generally such as in infinite dimensions, a frame for $\mathcal{V}$ is defined as any set of vectors $\{\ket{f_k}\}$ in $\mathcal{V}$ that satisfy
\begin{equation} \label{eq:frame_bounds}
    C \, ||v||^2 \leq \sum_k |\braket{f_k|v}|^2 \leq D \, ||v||^2 \text{ for all } \ket{v} \in \mathcal{V}
\end{equation}
for some $0 < C \leq D < \infty$ \cite{casazza2012finite}, where $||v||^2 = \braket{v|v}$ by definition. Equation \ref{eq:frame_bounds} assumes that the frame vectors lie in a countable set but can additionally be extended to include continuous frames. In these notes we only consider the simplest case scenario of a finite number $M$ of frame vectors with $1 \leq k \leq M$. We additionally assume that $C$ and $D$ are set to form the tightest possible bounds, in which case they are typically referred to as the upper and lower frame bounds of $\{\ket{f_k}\}$, respectively. A tight frame is a frame whose frame bounds are equal, i.e., $C = D$. Unlike in finite dimensions, in infinite dimensions the requirement that an arbitrary set of vectors spans $\mathcal{V}$ is a necessary but not sufficient condition to satisfy Equation \ref{eq:frame_bounds}.

Throughout these notes, $\{\ket{f_k}, 1 \leq k \leq M\}$ will always be used to denote a frame for $\mathcal{V}$. The frame coefficients of a given vector $\ket{v} \in \mathcal{V}$ will be denoted by $\{a_k = \braket{f_k|v}, 1 \leq k \leq M\}$. The $\{a_k\}$ are assumed to be real for simplicity, but this is easily generalized. It will be useful notationally to define the $M$-dimensional vector $\vec{a} = [a_1, \dots, a_M]^T$, which can itself be viewed as an element of a vector space $\mathcal{W}$ over the real numbers, equipped with the standard inner product and Euclidean norm. $\mathcal{W}$ is sometimes referred to as the coefficient space and we adopt that terminology in these notes. In finite dimensions with $M$ frame vectors, $\mathcal{W}$ is always isomorphic to $\mathbbm{R}^M$. Given any two vectors $\vec{u} = [u_1, \dots, u_M]^T$ and $\vec{w} = [w_1, \dots, w_M]^T$ in $\mathcal{W}$, the standard inner product between them will be denoted by $\langle \vec{u}, \vec{w} \rangle = \sum_k u_k \, w_k$. The squared norm of an arbitrary vector $\vec{w} \in \mathcal{W}$ will be denoted by $||w||^2 = \langle\vec{w},\vec{w}\rangle$. This notation coincides with that used to denote the squared norm of a vector $\ket{v} \in \mathcal{V}$, namely, $||v||^2 = \braket{v|v}$. We nevertheless utilize the same notation for both since it will always be clear from context which inner product is being used. Finally, the range and nullspace of an arbitrary linear transformation $T$ from $\mathcal{V}$ to $\mathcal{W}$ or $\mathcal{W}$ to $\mathcal{V}$ will always be denoted by $\text{range}(T)$ and $\text{null}(T)$, respectively.

\subsection{Analysis and Synthesis Operators} \label{ssec:analysis_and_synthesis_operators}

Associated with any frame $\{\ket{f_k}\}$ for $\mathcal{V}$ are two linear transformations referred to as the analysis and synthesis operators of the frame \cite{casazza2012finite}. The analysis operator $A$ takes as its input any $\ket{v} \in \mathcal{V}$ and generates the vector $\vec{a} = [a_1, \dots, a_M]^T$ of frame coefficients where $a_k = \braket{f_k|v}$ for $1 \leq k \leq M$,
\begin{equation}
    \ket{v} \in \mathcal{V} \longrightarrow \vec{a} = A(v) = [a_1, \dots, a_M]^T \in \mathcal{W}.
\end{equation}
Thus, $A$ maps every element of $\mathcal{V}$ to a specific element of $\mathcal{W}$. Since the frame vectors span $\mathcal{V}$, $A$ has rank $N$ and $\text{range}(A)$ is an $N$-dimensional subspace of $\mathcal{W}$. Since $||A(v)||^2 = \langle A(v), A(v) \rangle = \sum_k |a_k|^2$, Equation \ref{eq:frame_bounds} implies that
\begin{equation} \label{eq:analysis_operator_bounds}
    C \, ||v||^2 \leq ||A(v)||^2 \leq D \, ||v||^2 \text{ for all } \ket{v} \in \mathcal{V}.
\end{equation}
If $\{\ket{f_k}\}$ is a tight frame with frame bound $C$, then $||A(v)||^2 = C \, ||v||^2$ for all $\ket{v} \in \mathcal{V}$.

The synthesis operator $F$ takes as its input any vector $\vec{w} \in \mathcal{W}$ and produces as its output a vector in $\mathcal{V}$ according to the relation
\begin{equation}
    \vec{w} = [w_1, \dots, w_M]^T \in \mathcal{W} \longrightarrow \ket{v} = F(\vec{w}) = \sum_{k=1}^M w_k \ket{f_k} \in \mathcal{V}.
\end{equation}
Note that the components $\{w_k\}$ of $\vec{w}$ need not have been obtained by applying $A$ to some $\ket{v} \in \mathcal{V}$. Indeed, if $\vec{w}$ lies outside of $\text{range}(A)$ then there is no $\ket{v} \in \mathcal{V}$ such that $\braket{f_k|v} = w_k$ for $1 \leq k \leq M$. The range of $F$ is equal to the span of the frame vectors, which by assumption is equal to $\mathcal{V}$. Thus, $F$ has rank $N$ implying that $\text{null}(F)$ is an $(M-N)$-dimensional subspace of $\mathcal{W}$. It is straightforward to show that $A$ and $F$ are adjoints of each other, which additionally implies that
%
\begin{equation} \label{eq:adjoint_thm}
    \text{null}(F) = \text{range}(A)^\perp.
\end{equation}
In Equation \ref{eq:adjoint_thm} the superscript $\perp$ denotes the orthogonal complement of a subspace.

Now let $\{\ket{f_k}\}$ be a frame for $\mathcal{V}$ with analysis operator $A$ and let $\{\ket{\tilde{f}_k}\}$ be a possibly different frame for $\mathcal{V}$ with synthesis operator $\tilde{F}$. The composition of $\tilde{F}$ with $A$ is defined by
\begin{equation} \label{eq:analysis_synthesis_concatenation}
    \ket{v} \in \mathcal{V} \longrightarrow (\tilde{F} \circ A)(v) = \tilde{F}(A(v)) = \sum_{k=1}^M \braket{f_k|v}\ket{\tilde{f}_k}.
\end{equation}
In general, the vector $\tilde{F}(A(v))$ is different from $\ket{v}$. But when $\tilde{F}(A(v)) = \ket{v}$ for all $\ket{v} \in \mathcal{V}$, the two frames $\{\ket{f_k}\}$ and $\{\ket{\tilde{f}_k}\}$ are said to be dual to each other. Said differently, the two frames are dual to each other when $\tilde{F}$ is a left-inverse of $A$ or equivalently $A$ is a right-inverse of $\tilde{F}$. Dual frames are discussed in more detail next in Section \ref{ssec:dual_frames}. 
%
%
%
%

\begin{figure}
    \centering
    \includegraphics[width=4in]{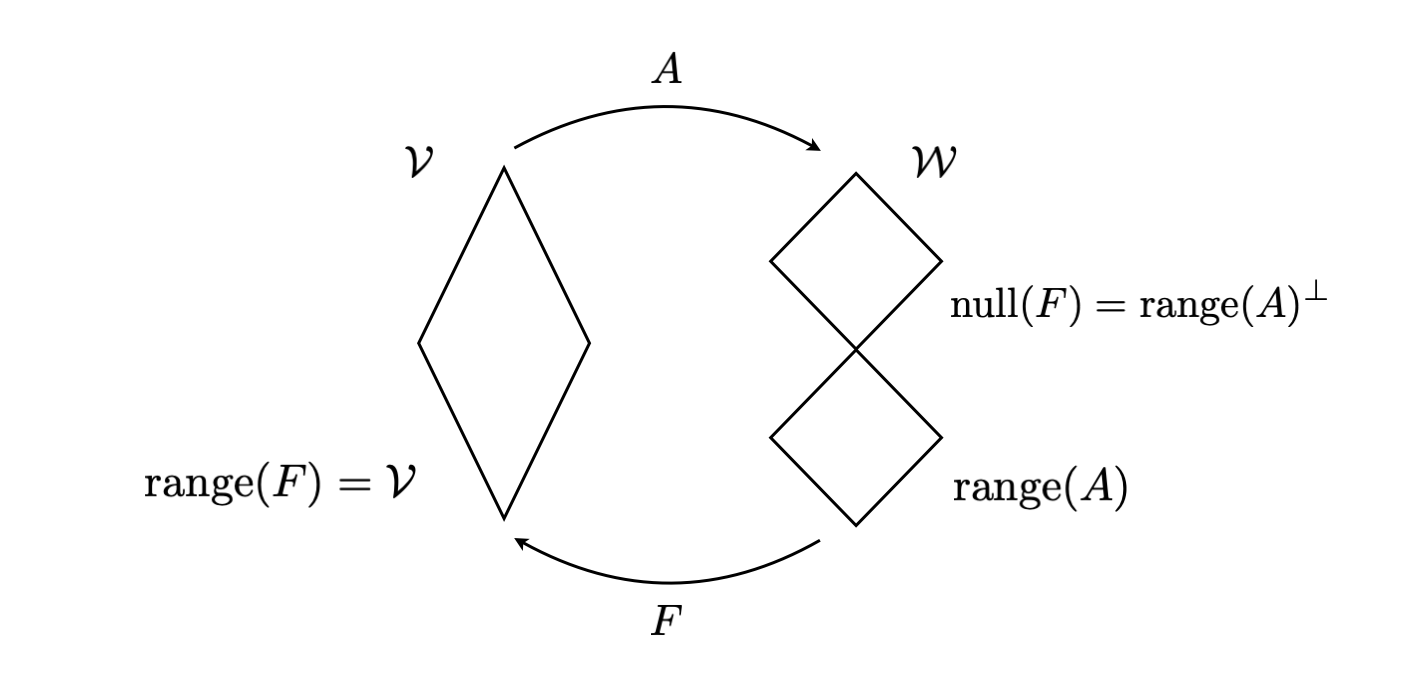}
    \caption{The analysis operator $A$ takes vectors in $\mathcal{V}$ to a subspace of $\mathcal{W}$ with the same dimension as $\mathcal{V}$. The synthesis operator $F$ takes coefficient vectors in $\mathcal{W}$ to vectors in $\mathcal{V}$.}
    \label{fig:adjoint_thms}
\end{figure}

\subsection{Dual Frames} \label{ssec:dual_frames}

Assume that $\{\ket{f_k}\}$ is a frame for $\mathcal{V}$ with analysis operator $A$ and that $\{\ket{\tilde{f}_k}\}$ is a possibly different frame for $\mathcal{V}$ with synthesis operator $\tilde{F}$. $\{\ket{\tilde{f}_k}\}$ is referred to as a dual frame of $\{\ket{f_k}\}$ if
\begin{equation} \label{eq:dual_frame_condition_with_maps}
    \ket{v} = (\tilde{F} \circ A)(v) = \tilde{F}(A(v)) \text{ for all } \ket{v} \in \mathcal{V}.
\end{equation}
It is straightforward to verify that if $\{\ket{\tilde{f}_k}\}$ is a dual frame of $\{\ket{f_k}\}$, then $\{\ket{f_k}\}$ is a dual frame of $\{\ket{\tilde{f}_k}$.
%

Given a frame $\{\ket{f_k}\}$ for $\mathcal{V}$, the dual frame of $\{\ket{f_k}\}$ is only unique when the frame vectors are linearly independent (in which case they form a basis). When the frame vectors are linearly dependent, one way of characterizing the set of all dual frames is to utilize the fact that every left-inverse of $A$ can be identified as the synthesis operator of a specific dual frame. As is clear from Equation \ref{eq:dual_frame_condition_with_maps}, the converse is also true -- the synthesis operator of any dual frame is always a left-inverse of $A$. If $L$ is any left-inverse of $A$, then by definition $L$ takes vectors in $\mathcal{W}$ to vectors in $\mathcal{V}$ in a way that satisfies
\begin{equation} \label{eq:left_inverse_defn}
    \ket{v} = L(A(v)) \text{ for all } \ket{v} \in \mathcal{V}.
\end{equation}
Equation \ref{eq:left_inverse_defn} implies that the image of $\text{range}(A)$ under $L$ is equal to $\mathcal{V}$, so $L$ has rank $N$. The nullspace of $L$ is thus an $(M-N)$-dimensional subspace of $\mathcal{W}$ that is linearly independent of $\text{range}(A)$. Distinct dual frames can be characterized by the distinct nullspaces of their synthesis operators. The dual frame whose synthesis operator $\tilde{F}$ is the left-inverse of $A$ with the property that $\text{null}(\tilde{F}) = \text{range}(A)^\perp$ is referred to as the canonical dual frame. The canonical dual frame is utilized in Section \ref{ssec:robustness_of_frame_representations} and discussed further in Appendix \ref{sec:optimality_of_canonical_dual}. From this point forward we will use $F_\text{can}$ to denote the synthesis operator of the canoncial dual frame of $\{\ket{f_k}\}$, so
\begin{equation} \label{eq:nullspace_of_F_can}
    \text{null}(F_\text{can}) = \text{range}(A)^\perp.
\end{equation}
Like the operator $F_\text{can}$, the synthesis operator $F$ of the original frame $\{\ket{f_k}\}$ also has the property that $\text{null}(F) = \text{range}(A)^\perp$, as stated in Equation \ref{eq:adjoint_thm}. However, it is important to note that $F_\text{can}$ is a left-inverse of $A$ by definition whereas $F$ is in general not a left-inverse of $A$. The exception is the case where $\{\ket{f_k}\}$ is a tight frame with frame bound $C$. In this scenario the canonical dual frame is the scaled frame $\{\ket{\tilde{f}_k} = \ket{f_k}/C\}$ and the corresponding synthesis operator is $F_\text{can} = F/C$. Thus, $F$ is a left-inverse of $A$ up to a constant factor. It will be useful in Section \ref{ssec:robustness_of_frame_representations} to note that when $\{\ket{f_k}\}$ is a tight frame with frame bound $C$, we have
\begin{equation} \label{eq:tight_frame_synthesis_reduces_norm}
    ||F_\text{can}(\vec{w})||^2 = \frac{||w||^2}{C} \text{ for all } \vec{w} \in \text{range}(A) \subset \mathcal{W}.
\end{equation}
To see why this is true, note that if $\vec{w}$ is an arbitrary vector in $\text{range}(A)$, then by definition there is some $\ket{v} \in \mathcal{V}$ such that $\vec{w} = A(v)$. Since $\{\ket{\tilde{f}_k}\}$ is a dual frame of $\{\ket{f_k}\}$, we have $F_\text{can}(\vec{w}) = F_\text{can}(A(v)) = \ket{v}$ which implies that $||F_\text{can}(\vec{w})||^2 = ||v||^2$. The  application of Equation \ref{eq:analysis_operator_bounds} to tight frames then leads to the relation $||w||^2/C = ||A(v)||^2 /C = ||v||^2$, and this implies that $||F_\text{can}(\vec{w})||^2 = ||w||^2 / C$.

\subsection{Robustness of Frame Representations} \label{ssec:robustness_of_frame_representations}

We now consider the problem of linearly reconstructing an arbitrary vector in $\mathcal{V}$ starting with imprecise versions of its frame coefficients. In Section \ref{ssec:application_to_estimation}, we describe how this scenario arises in the context of quantum state estimation. Given an unknown vector $\ket{v} \in \mathcal{V}$ and an analysis frame $\{\ket{f_k}\}$ for $\mathcal{V}$, $\ket{v}$ can always be written as
\begin{equation} \label{eq:v_decomposition_no_error}
    \ket{v} = \sum_{k=1}^M a_k \ket{\tilde{f}_k},
\end{equation}
where $\{\ket{\tilde{f}_k}\}$ is any dual frame of $\{\ket{f_k}\}$ and $a_k = \braket{f_k|v}$ for $1 \leq k \leq M$. Throughout Section \ref{ssec:robustness_of_frame_representations} we will denote the analysis and synthesis operators of $\{\ket{f_k}\}$ by $A$ and $F$, respectively, and the synthesis operator of $\{\ket{\tilde{f}_k}\}$ by $\tilde{F}$. The synthesis operator of the canonical dual frame of $\{\ket{f_k}\}$ will continue to be denoted by $F_\text{can}$. An important problem in classical signal processing is that of reconstructing $\ket{v}$ given only imprecise versions of the $\{a_k\}$ after they have been affected by some source of error. Below we describe a version of this problem that incorporates a specific model for the error source in more detail. The key takeaway is that when $\{\ket{f_k}\}$ is a certain type of tight frame and the error values are additive and uncorrelated, there is a tradeoff in the quality of reconstruction between the variance of the error values and the number $M$ of frame vectors.

\subsubsection{Problem Description and Solution} \label{sssec:problem_description}

We assume that the observed coefficients are $\{a_k + e_k\}$, where the individual error values $\{e_k\}$ have zero mean, variance $\sigma^2$, and collectively are pairwise uncorrelated. That is,
\begin{subequations} \label{eq:ek_assumptions}
\begin{alignat}{1}
    \mathbbm{E}[e_k] &= 0, \quad 1 \leq k \leq M, \\[5pt]
    \mathbbm{E}[e_j e_k] &= \begin{cases} \Delta^2 \text{ if } j = k \\ 0 \text{ if } j \neq k \end{cases}, \quad 1 \leq j, k \leq M. \label{eq:ek_assumptions:b}
\end{alignat}
\end{subequations}
Equations \ref{eq:ek_assumptions} have been shown to be a useful model mathematically in certain scenarios,  despite not always being literally true in practice -- see, for example, Chapter 4 of \cite{oppenheim2015signals}. The observed coefficient vector can be written as the sum of the true coefficient vector $\vec{a} = A(v)$ with the vector $\vec{e} = [e_1, \dots, e_M]^T \in \mathcal{W}$. For a given synthesis frame $\{\ket{\tilde{f}_k}\}$, the reconstructed vector $\ket{\hat{v}}$ is obtained by applying $\tilde{F}$ to the observed coefficient vector,
\begin{equation} \label{eq:vhat_defn}
    \ket{\hat{v}} = \tilde{F} \left( \, \vec{a} + \vec{e} \, \right) = \ket{v} + \ket{v_e}.
\end{equation}
In Equation \ref{eq:vhat_defn} we have defined the final error vector $\ket{v_e} = \tilde{F}(\,\vec{e}\,) = \sum_k e_k \ket{\tilde{f}_k}$. The objective is to find the synthesis frame that minimizes the expected value of the squared norm of $\ket{v_e}$, i.e., we want to minimize $\mathcal{E}$ where
\begin{equation} \label{eq:mathcal_E}
    \mathcal{E} = \mathbbm{E}\left[ ||v_e||^2 \right] =   \mathbbm{E}\left[ ||\tilde{F}(e)||^2 \right].
\end{equation}
It is well-known that as long as the error values are uncorrelated, the optimal synthesis frame that minimizes $\mathcal{E}$ is the canonical dual of the analysis frame \cite{goyal1998quantized}. This is true even if each of the $\{e_k\}$ have possibly different variances denoted by $\{\Delta_k^2\}$. The details of the derivation can be found in Appendix \ref{sec:optimality_of_canonical_dual}. The underlying idea is that the nullspace of the synthesis operator of the canonical dual frame contains the largest portion of the error vector $\vec{e}$ as compared to other dual frames.

\subsubsection{Application to Equal-Norm Tight Frames (ENTFs)} \label{sssec:application_to_entfs}

In these notes we will be particularly interested in the case where $\{\ket{f_k}\}$ is a tight frame for $\mathcal{V}$ with frame bound $C$, with the additional property that all of the frame vectors have the same norm, denoted by $a$. Such a frame is typically referred to as an equal norm tight frame (ENTF) \cite{casazza2012finite,casazza2003equal}. Mathematically, we have
\begin{subequations} \label{eq:entf_assumption}
\begin{alignat}{1}
    \sum_{k=1}^M |\braket{f_k|v}|^2 &= C \, ||v||^2 \text{ for all } \ket{v} \in \mathcal{V}, \\[5pt]
    ||f_k|| &= a, \quad 1 \leq k \leq M.
\end{alignat}
\end{subequations}
ENTFs are utilized, for example, in the context of oversampling in classical signal processing to reduce the effect of quantization noise on a bandlimited signal. They are also of interest in the quantum physics community in the form of tight IC POVMs as used for quantum state estimation. It can be shown \cite{casazza2003equal} that for an ENTF the following relationship holds,
\begin{equation} \label{eq:CN_equals_Ma2}
    C \, N = M \, a^2.
\end{equation}
Recall that $N$ denotes the dimension of $\mathcal{V}$. The canonical dual of an ENTF $\{\ket{f_k}\}$ is $\{\ket{\tilde{f}_k} = \ket{f_k}/C\}$. The corresponding synthesis operator is $\tilde{F} = F_\text{can} = F/C$. To find the minimum value of $\mathcal{E}$ in this scenario, we apply $F_\text{can}$ to the error vector $\vec{e}$ and evaluate the expected value of $||F_\text{can}(\,\vec{e}\,)||^2$. We first note that as stated in Section \ref{ssec:dual_frames} we have $\text{null}(F_\text{can}) = \text{range}(A)^\perp$. Consider writing the error vector as $\vec{e} = \vec{e}_1 + \vec{e}_2$ where $\vec{e}_1 \in \text{range}(A)$ and $\vec{e}_2 \in \text{range}(A)^\perp$. Then
\begin{equation}
    F_\text{can}(\,\vec{e}\,) = F_\text{can}(\vec{e}_1) + F_\text{can}(\vec{e}_2) = F_\text{can}(\vec{e}_1).
\end{equation}
Equation \ref{eq:tight_frame_synthesis_reduces_norm} then implies that for a given error vector,
\begin{equation}
    ||F_\text{can}(\,\vec{e}\,)||^2 = ||F_\text{can}(\vec{e}_1)||^2 = \frac{||\vec{e}_1||^2}{C}.
\end{equation}
Using the fact that the individual error values $\{e_k\}$ satisfy Equations \ref{eq:ek_assumptions}, it is straightforward to show that $\mathbbm{E}[||\vec{e}_1||^2] = N \Delta^2$. The minimum value $\mathcal{E}^*$ of $\mathcal{E}$ is thus
\begin{equation} \label{eq:snr_oversampling}
\mathcal{E}^* = \mathbbm{E}\left[||F_\text{can}(\,\vec{e}\,)||^2\right] = \frac{N \, \Delta^2}{C} = \frac{N^2 \, \Delta^2}{M \, a^2}.
\end{equation}
When the variances of the $\{e_k\}$ are not assumed to be identical for all values of $k$, it is straightforward to show that $\mathbbm{E}[||\vec{e}_1||^2] = (N/M) \sum_k \Delta_k^2$, so Equation \ref{eq:snr_oversampling} becomes
\begin{equation} \label{eq:snr_oversampling_diff_variances}
    \mathcal{E}^* = \frac{N}{MC} \sum_{k=1}^M \Delta_k^2 = \frac{N^2}{M^2 a^2} \sum_{k=1}^M \Delta_k^2.
\end{equation}
As expected, when $\Delta_k^2 = \Delta^2$ for all $1 \leq k \leq M$ Equation \ref{eq:snr_oversampling_diff_variances} reduces to Equation \ref{eq:snr_oversampling}.
%

\section{Frame Representations of Operator Spaces} \label{sec:operator_spaces}

The goal of Section \ref{sec:operator_spaces} is to extend the discussion in Section \ref{sec:frame_representations} to vector spaces whose elements are operators rather than vectors. We refer to operator-valued vector spaces as operator spaces for brevity. Our main focus throughout Sections \ref{ssec:defn_of_operator_valued_frame} to \ref{ssec:operators_with_constant_trace} is on operator spaces whose elements are Hermitian operators acting on a given vector-valued vector space. In Section \ref{ssec:application_to_d_equals_2} we give a concrete example in low dimensions.

We start by defining a certain operator space $\mathcal{V}$ along with a corresponding inner product. Given a vector-valued Hilbert space $\mathcal{H}$ of dimension $d$, the set of all linear operators from $\mathcal{H}$ to itself forms an operator space over the complex numbers. We define $\mathcal{V}$ to be the subspace of this larger operator space that contains all Hermitian operators on $\mathcal{H}$,
\begin{equation}
    \mathcal{V} = \left\{\text{Hermitian operators acting on } \mathcal{H}\right\}.
\end{equation}
$\mathcal{V}$ is an operator space over the real numbers but not the complex numbers since a complex multiple of a Hermitian operator is not guaranteed to be Hermitian. It is straightforward to show that $\mathcal{V}$ has dimension $N = d^2$.

Following a combination of the conventions in \cite{scott2006tight} and \cite{d2007optimal}, the inner product between any two operators $V_1, V_2 \in \mathcal{V}$ will be denoted using modified bra-ket notation as $\bbrakket{V_1}{V_2}$ and defined according to the relation
\begin{equation} \label{eq:inner_product_on_V}
    \bbrakket{V_1}{V_2} = \sum_{i=1}^d \bra{e_i} V_1 V_2 \ket{e_i} = \text{tr}(V_1V_2).
\end{equation}
In Equation \ref{eq:inner_product_on_V}, $\{\ket{e_i}\}$ is a fixed but arbitrary ONB for $\mathcal{H}$. Indeed, it is well-known that the trace of an operator when computed as the sum in Equation \ref{eq:inner_product_on_V} is independent of the ONB used. We further note that the function defined in Equation \ref{eq:inner_product_on_V} is a special case of the well-known Hilbert-Schmidt inner product \cite{scott2006tight}. It inherently depends on the vector inner product (denoted using traditional bra-ket notation) on $\mathcal{H}$.

\subsection{Definition of an Operator-Valued Frame} \label{ssec:defn_of_operator_valued_frame}

We repeat the definition of a frame for clarity using operator space notation. Any set of operators $\{F_k, 1 \leq k \leq M\}$ that lie in and span $\mathcal{V}$ form a frame for $\mathcal{V}$. More generally such as in infinite dimensions, a frame for $\mathcal{V}$ is any set of operators $\{F_k\}$ that lie in $\mathcal{V}$ and satisfy
\begin{equation} \label{eq:operator_valued_frame_bounds}
    C \, ||V||^2 \leq \sum_k |\bbrakket{F_k}{V}|^2 \leq D \, ||V||^2 \text{ for all } V \in \mathcal{V},
\end{equation}
for some $0 < C \leq D < \infty$ \cite{scott2006tight}. Equation \ref{eq:operator_valued_frame_bounds} assumes that the frame vectors lie in a finite or countably infinite set but can additionally be extended to include continuous frames. Again, we only consider the scenario in which $1 \leq k \leq M$. We will always assume that the values of $C$ and $D$ are set to form the tightest possible bounds, in which case they are referred to as the frame bounds of $\{F_k\}$. A tight frame for $\mathcal{V}$ is one whose frame bounds are equal.

Regardless of whether the number of frame vectors is finite or infinite, the definition of an operator frame given in Equation \ref{eq:operator_valued_frame_bounds} may also be generalized to the notion of a generalized operator frame with respect to a given measure \cite{scott2006tight}. In the terminology of \cite{scott2006tight}, a set of operators satisfying Equation \ref{eq:operator_valued_frame_bounds} is referred to as a generalized operator frame with respect to the counting measure. For brevity we do not discuss this or any other generalization further.

\subsection{Operators with Constant Trace} \label{ssec:operators_with_constant_trace}

In the context of quantum mechanics where $\mathcal{H}$ represents the state space of a quantum system, it will be useful to consider sets of operators in $\mathcal{V}$ that have constant trace. All density operators associated with a given state space $\mathcal{H}$ are elements of $\mathcal{V}$ with trace 1. For an arbitrary constant $\tau$, one way of categorizing the set of operators with trace $\tau$ relies on the decomposition of $\mathcal{V}$ into the following two orthogonal subspaces,
\begin{subequations}
\begin{alignat}{1}
    \mathcal{R} &= \{\text{traceless Hermitian operators on $\mathcal{H}$}\} \\[5pt]
    \mathcal{R}^\perp &= \text{span}\{I\},
\end{alignat}
\end{subequations}
where $I$ is the identity operator on $\mathcal{H}$. That the elements of $\mathcal{R}$ and $\mathcal{R}^\perp$ are indeed orthogonal to each other can be seen by taking the inner product of an arbitrary operator $V \in \mathcal{V}$ with $I$: $\bbrakket{I}{V} = \text{tr}(IV) = \text{tr}(V)$, which is equal to 0 if and only if $V$ has trace 0. The subspace $\mathcal{R}^\perp$ has dimension 1, while the subspace $\mathcal{R}$ has dimension $(N-1) = (d^2-1)$ and is always isomorphic to $\mathbbm{R}^{d^2-1}$ \cite{scott2006tight}.

Given an operator $V \in \mathcal{V}$, $V$ can always be written as the sum of its component in $\mathcal{R}$ and its component in $\mathcal{R}^\perp$. The latter component is equal to the orthogonal projection of $V$ onto $\mathcal{R}^\perp$, which can be expressed as
\begin{equation} \label{eq:P_Rperp_of_V}
    \mathcal{P}_{\mathcal{R}^\perp}(V) = \bbrakket{I}{V} \, I = \text{tr}(IV) \, I = \text{tr}(V) \, I.
\end{equation}
Thus, the set of all operators $V \in \mathcal{V}$ that have trace $\tau$ are those operators whose projection onto the direction of the identity is $\tau \, I$. These operators form a hyperplane in $\mathcal{V}$ that is orthogonal to the identity.

\subsection{Example with $d=2$} \label{ssec:application_to_d_equals_2}

We now explicitly describe the operator space $\mathcal{V}$ and its orthogonal subspaces $\mathcal{R}$ and $\mathcal{R}^\perp$ when $\mathcal{H} = \mathbbm{C}^2$. Our intent aside from providing a concrete example in low dimensions is to hopefully also present some amount of geometric intuition regarding where operators with constant trace and also positive semidefinite operators lie in $\mathcal{V}$. While none of the concepts presented in Section \ref{ssec:application_to_d_equals_2} are specific to the context of quantum mechanics, they are relevant to the simulations presented in Section \ref{sec:application_to_quantum} involving qubit systems since by the state space of a qubit is always isomorphic to $\mathbbm{C}^2$.

When $\mathcal{H} = \mathbbm{C}^2$, $\mathcal{V}$ has dimension $d^2 = 4$. It is well-known that the following operators form an orthonormal basis for $\mathcal{V}$ under the inner product defined by Equation \ref{eq:inner_product_on_V},
\begin{equation} \label{eq:pauli_basis}
    \{\hat{\sigma}_0, \, \hat{\sigma}_1, \, \hat{\sigma}_2, \, \hat{\sigma}_3\} = \left\{ \frac{I}{\sqrt{2}}, \, \frac{\sigma_1}{\sqrt{2}}, \, \frac{\sigma_2}{\sqrt{2}}, \, \frac{\sigma_3}{\sqrt{2}} \right\}.
\end{equation}
In Equation \ref{eq:pauli_basis}, $\sigma_0 = I$ is defined for convenience, $\{\sigma_1, \sigma_2, \sigma_3\}$ are the Pauli operators, and the symbol \^{} is used to denote multiplication by $2^{-1/2}$. We have $\mathcal{R}^\perp = \text{span}\{\hat{\sigma}_0\} = \text{span}\{\sigma_0\}$ and $\mathcal{R} = \text{span}\{\sigma_1, \sigma_2, \sigma_3\}$. Obviously, the choice of orthonormal basis for $\mathcal{R}$ is not unique. However, the Pauli operators will prove to be a convenient choice in the context of quantum mechanics as they are directly related to the well-known representation of an arbitrary qubit density operator in terms of its Bloch vector.

Given a Hermitian operator $V$ acting on $\mathcal{H}$, $V$ can always be written as a linear combination of the $\{\hat{\sigma}_i, 0 \leq i \leq 3\}$,
\begin{equation} \label{eq:V_in_terms_of_the_sigmas}
    V = \sum_{i=0}^3 \bbrakket{\hat{\sigma}_i}{V} \, \hat{\sigma}_i = \sum_{i=0}^3 c_i \, \hat{\sigma}_i,
\end{equation}
where $c_i = \bbrakket{\hat{\sigma}_i}{V} = \text{tr}(\hat{\sigma}_i V)$ for $0 \leq i \leq 3$. It is straightforward to verify that
\begin{subequations} \label{eq:trace_and_pos_semidef}
\begin{alignat}{1}
    \text{tr}(V) = \tau \, &\Leftrightarrow \, c_0 = \frac{\tau}{\sqrt{2}}, \label{eq:trace_ci}\\[5pt]
    V \text{ positive semidefinite} \, &\Leftrightarrow \, c_0^2 \geq c_1^2 + c_2^2 + c_3^2, \label{eq:pos_semidef_ci}
\end{alignat}
\end{subequations}
where $\tau$ is an arbitrary constant. Equation \ref{eq:trace_ci} is essentially a restatement of Equation \ref{eq:P_Rperp_of_V} and implies that the set of all operators in $\mathcal{V}$ that have fixed trace $\tau$ forms a hyperplane in $\mathcal{V}$. Equation \ref{eq:pos_semidef_ci} can be verified by solving for the eigenvalues of $V$ in terms of the $\{c_i\}$ and setting them to be non-negative. It implies that the set of all positive semidefinite operators in $\mathcal{V}$lie on or within a fixed cone in $\mathcal{V}$. Both statements are generalizable to higher dimensions [Tight IC POVMs, Minimal Informationally Complete Measurements for Pure States] and are visualized in Example \ref{ex:hyperplane_and_cone} below.

\begin{example} \label{ex:hyperplane_and_cone} \normalfont
For the development of geometric intuition we consider analogous constraints to Equations \ref{eq:trace_and_pos_semidef} in 3 dimensions. We temporarily define $\mathcal{V} = \mathbbm{R}^3$ with dimension $N=3$ and ONB $\vec{b}_0 = [0, 0, 1]^T$, $\vec{b}_1 = [0, 1, 0]^T$, and $\vec{b}_2 = [1, 0, 0]^T$. An arbitrary vector $\vec{x} \in \mathbbm{R}^3$ can always be expressed as
\begin{equation} \label{eq:hyperplane_R3_example}
    \vec{x} = c_0 \, \vec{b}_0 + c_1 \, \vec{b}_1 + c_2 \, \vec{b}_2.
\end{equation}
In Equation \ref{eq:hyperplane_R3_example}, $\langle\cdot, \cdot\rangle$ denotes the standard dot product and $c_i = \langle\vec{b}_i, \vec{x}\rangle$ for $0 \leq i \leq 2$. As shown in Figure \ref{fig:hyperplane_intersection_with_cone}, the set of vectors in $\mathbbm{R}^3$ that satisfy $c_0 = 2^{-1/2}$ lie on a hyperplane while the set of vectors that satisfy $c_1^2 + c_2^2 \leq c_0^2$ lie on or within a cone. The set of vectors that satisfy both of the constraints lies at the intersection of the hyperplane and the cone and takes the form of an $(N-1) = 2$ dimensional ball (i.e., a circle).

\begin{figure}
    \centering
    \includegraphics[height=2.5in]{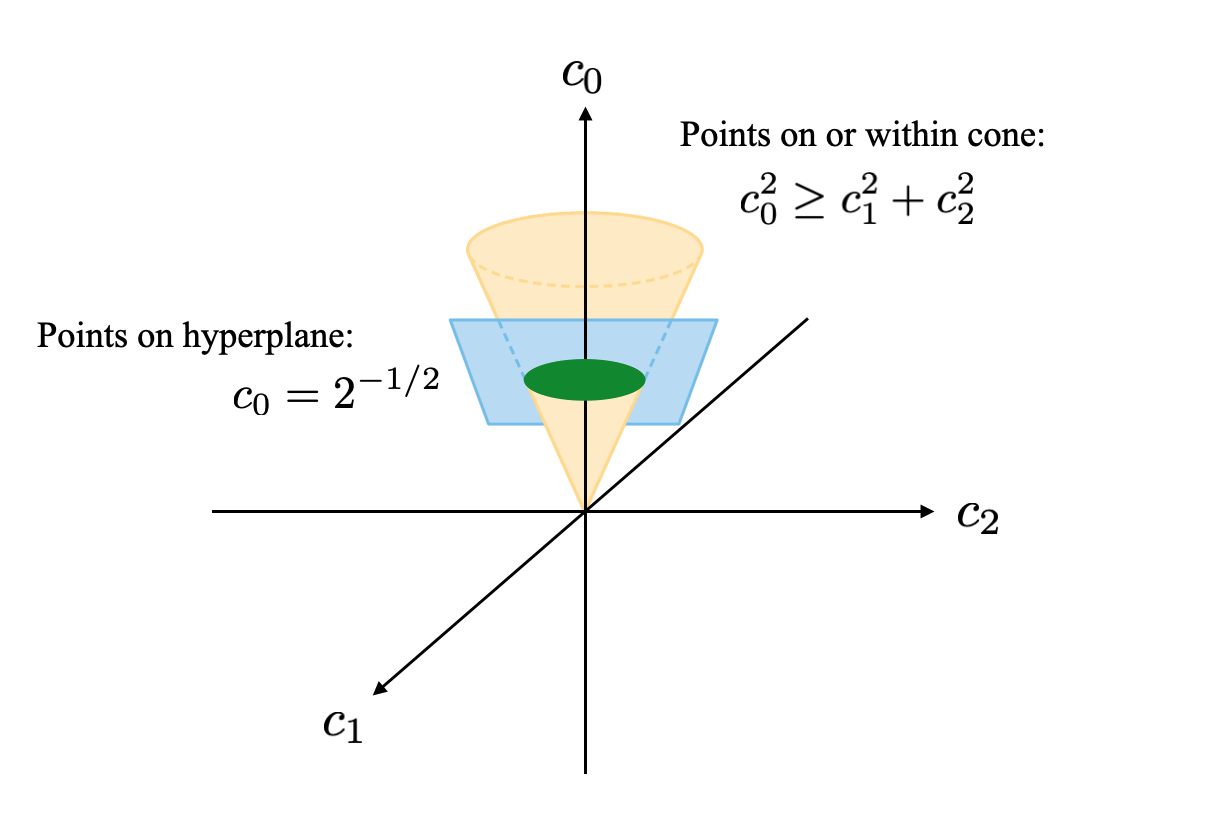}
    \caption{Illustration of the constraints described in Example \ref{ex:hyperplane_and_cone}. Figure adapted from \cite{internetfig}.}
    \label{fig:hyperplane_intersection_with_cone}
\end{figure}

\end{example}

\section{Operator Spaces in Quantum Mechanics} \label{sec:operator_spaces_in_quantum_mechanics}

The consideration of operator spaces and the surrounding mathematical framework in the context of quantum mechanics is an essential tool in the study of informationally-complete (IC) POVMs. POVMs are collections of Hermitian operators used in quantum mechanics as part of a mathematical description of the process of quantum measurement. When a quantum system in a given state is measured, the POVM associated with the measurement being performed can be used to specify a probability distribution over the possible measurement outcomes. An IC POVM is a POVM for which distinct quantum states are always mapped to distinct probability distributions. This is important because it implies that an unknown quantum state can be reconstructed from its probability distribution.

Throughout Section \ref{sec:operator_spaces_in_quantum_mechanics}, $\mathcal{H}$ will always represent the state space of a quantum system with dimension $d$. As in Section \ref{sec:operator_spaces}, $\mathcal{V}$ will always be used to denote the operator space of all Hermitian operators acting on $\mathcal{H}$. We start by reviewing the quantum state and measurement postulates in Section \ref{ssec:postulates}. In Section \ref{ssec:density_operators_and_povm_elements_in_V} we apply the statements made in Section \ref{ssec:operators_with_constant_trace} to density operators and POVM elements. IC POVMs are defined and discussed in Section \ref{ssec:ic_povms}, followed by a discussion of a special class of POVMs referred to as tight IC POVMs in Section \ref{ssec:tight_ic_povms}.

\subsection{The Postulates of Quantum Mechanics} \label{ssec:postulates}

We summarize two of the postulates as stated in \cite{nielsen2016quantum} as they relate to this article. The others are not directly relevant to our discussion here and are omitted. The state of an isolated physical system can be represented by a density operator $\rho$ that acts on a complex Hilbert space $\mathcal{H}$. We assume for convenience that $\mathcal{H}$ is finite dimensional with dimension $d$. $\rho$ is always a non-negative Hermitian operator that has trace equal to 1. Thus, it can be written in terms of its eigenbasis as
\begin{equation}
    \rho = \sum_{i=1}^d \lambda_i \ket{x_i}\bra{x_i},
\end{equation}
where the $\{\ket{x_i}\}$ form an ONB for $\mathcal{H}$ and the $\{\lambda_i\}$ are real and satisfy $0 \leq \lambda_i \leq 1$, $\sum_i \lambda_i = 1$.

Quantum measurements are described by a collection $\{A_k\}$ of measurement elements that are by definition Hermitian operators acting on $\mathcal{H}$. Each measurement element $A_k$ corresponds to a different possible measurement outcome. We will assume for simplicity that $1 \leq k \leq M$.\footnote{the usage of the index $k$ with range $1 \leq k \leq M$ coincides with our choice of indexing for frame vectors $\{\ket{f_k}\}$. This is intentional since we will eventually associate each POVM element $E_k = A_k^\dagger A_k$ with a frame vector of $\mathcal{V}$, as explained in Section \ref{ssec:ic_povms}.} Measurement elements always satisfy a completeness relation on $\mathcal{H}$,
\begin{equation} \label{eq:quantum_measurement_completeness}
    \sum_{k=1}^M A_k^\dagger A_k = I.
\end{equation}
If the state of a quantum system is described by the density operator $\rho = \sum_i \lambda_i \ket{x_i}\bra{x_i}$ immediately before a measurement with elements $\{A_k\}$, then with probability
\begin{equation} \label{eq:pm_untraced}
    p(k) = \sum_{i=1}^d \lambda_i \bra{x_i}A_k^\dagger A_k\ket{x_i}
\end{equation}
the $k$th measurement outcome occurs. Once observed, the $k$th measurement outcome indicates that the state of the system has collapsed to the $k$th post-measurement state, denoted by $\rho_k$. The value of $\rho_k$ can be specified in terms of $\rho$ and $A_k$, but since the exact expression is not relevant to these notes it is omitted.

A given quantum measurement with measurement elements $\{A_k\}$ has an associated set of operators $\{E_k = A_k^\dagger A_k\}$ that form a POVM, i.e., a set of positive-semidefinite, Hermitian operators acting on $\mathcal{H}$ that sum to the identity \cite{berberian1966notes}. In terms of the $\{E_k\}$, Equation \ref{eq:pm_untraced} can be written as

\begin{equation} \label{eq:pk_defn_2}
    p(k) = \sum_{i=1}^d \lambda_i \bra{x_i}E_k\ket{x_i}.
\end{equation}
%

\subsection{Density Operators and POVM Elements in $\mathcal{V}$} \label{ssec:density_operators_and_povm_elements_in_V}

Since density operators and POVM elements are by definition Hermitian operators acting on $\mathcal{H}$, they are also elements of $\mathcal{V}$. An important observation is that in terms of the inner product defined in Equation \ref{eq:inner_product_on_V}, the outcome probabilities $\{p(k)\}$ in Equation \ref{eq:pk_defn_2} can be expressed as
\begin{equation} \label{eq:pk_bbrakket}
    p(k) = \bbrakket{E_k}{\rho} = \text{tr}(E_k\rho).
\end{equation}
The expression of the $\{p(k)\}$ as the inner products of the $\{E_k\}$ with the density operator $\rho$ is a key concept underlying the connection between IC POVMs and frames for $\mathcal{V}$.

It will also be useful to describe density operators and POVM elements in relation to the subspaces $\mathcal{R}$ and $\mathcal{R}^\perp$. All density operators must have trace 1 by definition, so as explained in Section \ref{ssec:operators_with_constant_trace} they lie in a hyperplane in $\mathcal{V}$ that is orthogonal to the identity. An equivalent statement is that given an arbitrary density operator $\rho$, the shifted operator $(\rho - I/d)$ is always an element of $\mathcal{R}$. This is due to the fact that both $\rho$ and $I/d$ have trace 1, so their difference has trace 0. Since all density operators must also be positive semidefinite, they form only a subset of all elements of the hyperplane. Equivalently, the set of all elements of $\mathcal{V}$ that can be written in the form $(\rho-I/d)$ where $\rho$ is a valid density operator is only a subset of all elements of $\mathcal{R}$. 

Given an arbitrary POVM $\{E_k\}$, the scaled and shifted operators $\{S_k = E_k/\text{tr}(E_k) - I/d\}$ are always elements of $\mathcal{R}$. The $\{S_k\}$ are referred to in \cite{scott2006tight} as a positive operator-valued density. When the $\{S_k\}$ form what is referred to as a tight frame for $\mathcal{R}$ with respect to the trace measure, $\{E_k\}$ is referred to as a tight IC POVM. Tight IC POVMs are elaborated on further in Section \ref{ssec:tight_ic_povms}.

\begin{example} \normalfont
When $\mathcal{H}$ is the state of a qubit, the analysis presented in Section \ref{ssec:application_to_d_equals_2} applies to $\mathcal{V}$. An arbitrary density operator $\rho$ can be written as a linear combination of the identity and the Pauli operators,
\begin{equation}
    \rho = \sum_{i=0}^3 c_i \, \hat{\sigma}_i,
\end{equation}
where $c_i = \bbrakket{\hat{\sigma}_i}{\rho} = \text{tr}(\hat{\sigma}_i\rho)$ for $0 \leq i \leq 3$. Since $\rho$ must have trace 1 and be positive semidefinite, according to Equations \ref{eq:trace_and_pos_semidef} we must have $c_0 = 2^{-1/2}$ and $c_1^2 + c_2^2 + c_3^2 \leq c_0^2 = 1/2$. The set of all valid density operators therefore lies at the intersection of the hyperplane defined by the constraint $c_0 = 2^{-1/2}$ with the cone defined by the constratin $c_1^2 + c_2^2 + c_3^2 \leq c_0^2 = 1/2$. The intersection takes the form of a ball in $(d^2-1) = 3$ dimensions with radius $2^{-1/2}$. To within a constant factor, the ball corresponds to the well-known Bloch ball (which has radius 1) and the coefficients $(c_1, c_2, c_3)$ correspond to the Bloch vector of $\rho$.
\end{example}

\begin{example} \label{ex:povms_qubit} \normalfont
Again assume that $\mathcal{H}$ represents the state space of a qubit and let $\{E_k\}$ be a POVM whose elements can be expressed as
\begin{equation} \label{eq:Ek_qubit_onb_expansion}
    E_k = \sum_{i=0}^3 c_{ki} \, \hat{\sigma}_i, \quad 1 \leq k \leq M.
\end{equation}
In Equation \ref{eq:Ek_qubit_onb_expansion} we have again defined $c_{ki} = \bbrakket{\hat{\sigma}_i}{E_k}$ for $0 \leq i \leq 3$ and $1 \leq k \leq M$. The shifted and scaled operators $\{S_k = E_k/\text{tr}(E_k) - I/2\}$ are
\begin{equation} \label{eq:Pk_qubit_onb_expansion}
    S_k = 0 \cdot \hat{\sigma}_0 + \sqrt{2} \, \sum_{i=1}^3 \frac{c_{ki}}{c_{k0}} \, \hat{\sigma}_i, \quad 1 \leq k \leq M.
\end{equation}
Note that since the $\{S_k\}$ all have zero trace, their components in the direction of $\hat{\sigma}_0$ are all equal to zero. By definition, each of the $\{E_k\}$ must be positive semidefinite and thus must have non-negative trace. The additional requirement that $\sum_k E_k = I = \sqrt{2} \, \hat{\sigma}_0$ implies that the traces of the $\{E_k\}$ must sum to $\text{tr}(I) = d = 2$ and that their components in the $\hat{\sigma}_i$ direction for each $1 \leq i \leq 3$ must sum to zero. In summary, the $\{c_{ki}\}$ must satisfy
\begin{subequations} \label{eq:povm_element_ci}
\begin{alignat}{1}
    0 \leq c_{k0} &\leq d, \quad 1 \leq k \leq M \\[5pt]
    c_{k1}^2 + c_{k2}^2 + c_{k3}^2 &\leq c_{k0}^2, \quad 1 \leq k \leq M \\[5pt]
    \sum_{k=1}^M c_{k0} &= \frac{1}{\sqrt{2}}\sum_{k=1}^M \text{tr}(E_k) = \frac{d}{\sqrt{2}}, \\[5pt]
    \sum_{k=1}^M c_{k1} &= \sum_{k=1}^M c_{k2} = \sum_{k=1}^M c_{k3} = 0.
\end{alignat}
\end{subequations}
When the sets of coefficients $(c_{k1}, c_{k2}, c_{k3})$ for $1 \leq k \leq M$ correspond to the vertices of one of the five Platonic solids, it is typically said that the POVM was constructed from that Platonic solid \cite{huangjun2012quantum,zhu2014quantum,decker2004quantum,slomczynski2016highly} When an octahedron is used, the POVM is often described in the literature as having been constructed from three mutually unbiased bases, or MUBs, for the state space of the qubit. POVMs constructed from Platonic solids are used in Section \ref{sec:application_to_quantum}.
\end{example}

\subsection{Informationally Complete POVMs} \label{ssec:ic_povms}

An informationally complete (IC) POVM is one that maps each possible density operator to a unique sequence of probabilities. Given two density operators $\rho_1$ and $\rho_2$ as well as a POVM $\{E_k\}$, the corresponding probability distributions are $p_i(k) = \text{tr}(E_k \rho_i)$ for $i = 1,2$. If the POVM is IC then we have
\begin{equation}
    p_1(k) = p_2(k), \quad 1 \leq k \leq M,
\end{equation}
if and only if $\rho_1 = \rho_2$. A fundamental result regarding IC POVMs states that in finite dimensions, a given POVM is IC if and only if its elements are a frame for $\mathcal{V}$. In short, for a set of operators $\{E_k\}$ in $\mathcal{V}$,
\begin{equation} \label{eq:ic_povm_iff_frame}
    \{E_k\} \text{ is an IC POVM } \Leftrightarrow \{E_k\} \text{ is a valid POVM and } \{E_k\} \text{ is a frame for } \mathcal{V}.
\end{equation}
The statement in Equation \ref{eq:ic_povm_iff_frame} can be generalized to infinite dimensions and to more general definitions of operator frames \cite{scott2006tight}. The terms ``minimal IC POVM'' and ``informationally overcomplete (IOC) POVM'' are sometimes used to differentiate between those IC POVMs whose elements are linearly indpendent and thus form a basis for $\mathcal{V}$ and those whose elements are linearly dependent, respectively \cite{scott2006tight,caves2002unknown,flammia2005minimal,appleby2007symmetric,huangjun2012quantum,zhu2014quantum}. The result summarized by Equation \ref{eq:ic_povm_iff_frame} is important enough that we include one direction of the derivation below, in part to provide some intuition for why it is true. Roughly, the underlying idea is that if a POVM $\{E_k\}$ is a frame for $\mathcal{V}$, then every operator $V$ in $\mathcal{V}$ has a unique set of frame coefficients $\{a_k = \bbrakket{E_k}{V}\}$. When $V = \rho$ is a density operator, its frame coefficients are equal to the probabilities $\{p(k) = \bbrakket{E_k}{V}\}$, so every density operator is mapped to a unique set of probabilities. A derivation of the other direction of the result can be found in, for example, \cite{bodmann2017short,scott2006tight}.

Assume that a POVM $\{E_k\}$ forms a frame for $\mathcal{V}$ and let $\mathbf{A}$ be its analysis operator.\footnote{The analysis operator of a frame whose elements are themselves operators on a vector-valued vector space is a ``superoperator'', i.e., a linear operator acting on an operator-valued vector space. Superoperators will be denoted using boldfaced letters.} $\mathbf{A}$ maps every density operator $\rho \in \mathcal{V}$ to the vector in $\mathcal{W}$ whose elements are the probabilities $\{p(k) = \bbrakket{E_k}{\rho}\}$,
\begin{equation}
    \rho \in \mathcal{V} \longrightarrow \vec{a} = \mathbf{A}(V) = [\,p(1), \dots, p(M)\,]^T \in \mathcal{W}.
\end{equation}
To show that $\{E_k\}$ is IC, it is sufficient to show that if two density operators have the same probability sequences with respect to this POVM, then they must be identical. This is a direct consequence of the fact that $A$ is full-rank and therefore left-invertible as stated in Section \ref{ssec:analysis_and_synthesis_operators}. Specifically, since the $\{E_k\}$ span $\mathcal{V}$, no $V \in \mathcal{V}$ is orthogonal to all of them. Therefore, if $\mathbf{A}(V) = 0$ for some $V \in \mathcal{V}$ then we must have $V = 0$. Consider the action of $\mathbf{A}$ on two arbitrary density operators $\rho_1$, $\rho_2 \in \mathcal{V}$. We have
\begin{subequations}
\begin{alignat}{1}
    \mathbf{A}(\rho_1) &= [\,p_0(1), \dots, p_0(M)\,]^T \in \mathcal{W}, \\[5pt]
    \mathbf{A}(\rho_2) &= [\,p_1(1), \dots, p_1(M)\,]^T \in \mathcal{W},
\end{alignat}
\end{subequations}
where $p_i(k) = \bbrakket{E_k}{\rho_i}$ for $i = 1, 2$ and $1 \leq k \leq M$. If $p_1(k) = p_2(k)$ for $1 \leq k \leq M$, then $\mathbf{A}( \, \rho_1-\rho_2) = 0$ implying that $\rho_1-\rho_2 =0$, i.e., $\rho_1 = \rho_2$. The same is true for any two operators $V_1, V_2 \in \mathcal{V}$. If $\bbrakket{E_k}{V_1} = \bbrakket{E_k}{V_2}$ for all $1 \leq k \leq M$, then $V_1 = V_2$.

IC POVMs are commonly studied in the context of quantum state estimation \cite{huangjun2012quantum,zhu2014quantum,zhu2015super,vrehavcek2015determining,scott2006tight,adamson2010improving,renes2004symmetric,caves2002unknown}, in which the objective is to reconstruct an unknown density operator from its probability values stemming from a given POVM. Obviously, the ability to recover an arbitrary density operator using only the probability values requires the POVM to be IC. But even if an IC POVM is employed, exact recovery of the probability values can only be achieved if we are able to measure an infinitely large collection of systems, all prepared in the unknown state we wish to estimate. This is in general not possible in practice, and one motivation for using IOC POVMs is to mitigate the error caused by finite sample size estimations of the probabilities. This topic is also a main motivation for the simulations presented in Section \ref{sec:application_to_quantum}.

Another important issue in the use of IC POVMs to estimate unknown quantum states is that the reconstruction procedure implicitly requires computation of the dual frame of the POVM elements. This is in general a difficult task because it requires the inversion of a linear operator on $\mathcal{V}$, which is itself a ``superoperator'' \cite{scott2006tight}. Thus, IC POVMs whose duals are more easily computed are of great interest to the quantum physics community. Tight IC POVMs, defined next in Section \ref{ssec:tight_ic_povms}, are some of the most extensively studied and well-understood.

\subsection{Tight IC POVMs} \label{ssec:tight_ic_povms}

A tight IC POVM could be naturally defined as an IC POVM whose elements form a tight frame for $\mathcal{V}$. However, the definition is in fact slightly more nuanced as it takes into account the fact that all density operators lie within a hyperplane of $\mathcal{V}$. Briefly, the underlying logic is that when the hyperplane containing all density operators is shifted to the origin, it is identical to the subspace $\mathcal{R}$ of $\mathcal{V}$. The elements of a POVM may always be scaled and shifted to lie in $\mathcal{R}$, and when the scaled and shifted versions of the POVM elements form what is referred to as a tight frame for $\mathcal{R}$ with respect to the trace measure, the POVM is referred to as a tight IC POVM.

Recall from Section \ref{ssec:density_operators_and_povm_elements_in_V} that given an arbitrary density operator $\rho$, the shifted operator $(\rho - I/d)$ is always an element of $\mathcal{R}$. Given an arbitrary POVM $\{E_k\}$, the scaled and shifted operators defined by $\{S_k = E_k/\text{tr}(E_k) - I/d\}$ also lie in $\mathcal{R}$. In \cite{scott2006tight} a tight IC POVM was defined as a POVM for which the $\{S_k\}$ satisfy
\begin{equation} \label{eq:tight_ic_povm_trace_measure}
    \sum_{k=1}^M \text{tr}(E_k) \, |\bbrakket{S_k}{V}|^2 = C \, ||V||^2 \text{ for all } V \in \mathcal{R},
\end{equation}
for some constant $C>0$. Since all POVM elements must have non-negative trace, Equation \ref{eq:tight_ic_povm_trace_measure} may be re-written using the operators $\{Q_k = \sqrt{\text{tr}(E_k)} \, S_k\}$, resulting in the equivalent form
\begin{equation} \label{eq:tight_ic_povm}
    \sum_{k=1}^M |\bbrakket{Q_k}{V}|^2 = C \, ||V||^2 \text{ for all } V \in \mathcal{R}.
\end{equation}
Thus, in our terminology a tight IC POVM is a POVM for which the $\{Q_k\}$ form a tight frame for $\mathcal{R}$. Note that if the operators $\{Q_k\}$ associated with a given POVM $\{E_k\}$ satisfy Equation \ref{eq:tight_ic_povm}, then it is straightforward to show that the $\{E_k\}$ form a frame for $\mathcal{V}$ and thus that the POVM is IC. A well-known class of tight IC POVMs are those constructed from the five Platonic solids, which were mentioned in Example \ref{ex:povms_qubit}. Comparing Equation \ref{eq:tight_ic_povm_trace_measure} to the definition of an operator-valued frame in Equation \ref{eq:operator_valued_frame_bounds}, it is clear that the only difference (aside from the substitution of $\mathcal{R}$ for $\mathcal{V}$) is the extra factor of $\text{tr}(E_k)$ in each term of the sum. This factor is the reason that in the terminology of \cite{scott2006tight}, any set of operators $\{S_k\}$ satisfying Equation \ref{eq:tight_ic_povm_trace_measure} are said to form a tight frame for $\mathcal{R}$ with respect to the trace measure.

\section{Application to Quantum State Estimation and Binary Detection} \label{sec:application_to_quantum}

The focus of Section \ref{ssec:robustness_of_frame_representations} was the problem of linearly reconstructing an arbitrary element $\ket{v}$ of a vector-valued vector space $\mathcal{V}$ starting with imprecise versions of its frame coefficients $\{a_k = \braket{f_k|v}\}$. In Section \ref{ssec:application_to_estimation} we describe an analogous problem in the context of quantum state estimation. We provide evidence through simulation that the analysis of Section \ref{ssec:robustness_of_frame_representations} is a useful model in this context. In Section \ref{ssec:application_to_detection} we move to the problem of quantum binary detection. Given a collection of $L$ quantum systems all prepared in the same unknown state, the objective is to measure each system individually and with quantum measurements that have the same fixed POVM $\{E_k, 1 \leq k \leq M\}$. We demonstrate through simulations for qubit states that, at least for this formulation of the detection problem, there is a tradeoff in detection performance between the collection size $L$ and the number of POVM elements $M$.

\subsection{Quantum State Estimation} \label{ssec:application_to_estimation}

Let $\rho$ be an unknown density operator and let $\{E_k\}$ be an arbitrary tight IC POVM. We now consider a variation of the problem stated in Section \ref{ssec:robustness_of_frame_representations} in which the vector $\ket{v}$ lying in $\mathcal{V}$ is replaced by the shifted operator $(\rho-I/d)$ lying in the operator space $\mathcal{R}$. The analysis frame $\{\ket{f_k}\}$ is replaced by the operators $\{Q_k = E_k / \sqrt{\text{tr}(E_k)} - \sqrt{\text{tr}(E_k)} I/d\}$ that were defined in Section \ref{ssec:tight_ic_povms}. Since the $\{E_k\}$ are a tight IC POVM by assumption, the $\{Q_k\}$ form a tight frame for $\mathcal{R}$. We will denote the frame bound of the $\{Q_k\}$ by $C$ and will additionally assume that they all have norm $a$. Thus, the $\{Q_k\}$ are an ENTF for $\mathcal{R}$ satisfying
\begin{subequations} \label{eq:Qk_assumption}
\begin{alignat}{1}
    \sum_{k=1}^M |\bbrakket{Q_k}{V}|^2 &= C \, ||V||^2 \text{ for all } V \in \mathcal{R}, \\[5pt]
    ||Q_k|| &= a, \quad 1 \leq k \leq M.
\end{alignat}
\end{subequations}
In analogy with Equation \ref{eq:v_decomposition_no_error}, $(\rho-I/d)$ can always be expressed as
\begin{equation} \label{eq:rho_decomposition_no_error}
    \rho - \frac{I}{d} = \sum_{k=1}^M \bbrakket{Q_k}{\rho-I/d} \, \tilde{Q}_k = \sum_{k=1}^M a_k \, \tilde{Q}_k
\end{equation}
where $\{a_k = \bbrakket{Q_k}{\rho-I/d}\}$ and $\{\tilde{Q}_k\}$ is any dual frame of $\{Q_k\}$. In terms of the probabilities $\{p(k) = \bbrakket{E_k}{\rho}\}$, it is straightforward to show that the $\{a_k\}$ can be written as
\begin{equation} \label{eq:ak_in_terms_of_pk}
    a_k = \bbrakket{Q_k}{\rho-I/d} = \frac{p(k)}{\sqrt{\text{tr}(E_k)}} - \frac{\sqrt{\text{tr}(E_k)}}{d}.
\end{equation}
We assume that the observed, imprecise values $\{\hat{a}_k = a_k + e_k\}$ of the frame coefficients are obtained as follows. Given $L$ identically prepared quantum systems all in the state $\rho$, the true probabilities $\{p(k)\}$ are estimated by performing a quantum measurement with associated POVM $\{E_k\}$ on each system. The set of all measurement outcomes are used to compute estimates of the true probabilities in the form of the relative frequencies. If $\ell_k$ is the number of times the measurement outcome associated with $E_k$ occurred, the relative frequencies are $\{\hat{p}(k) = \ell_k/L\}$. They can always be expressed as $\{\hat{p}(k) = p(k) + d_k\}$ for a set of constants $\{d_k\}$. The $\{\hat{a}_k\}$ are obtained by replacing $p(k)$ with $\hat{p}(k)$ in Equation \ref{eq:ak_in_terms_of_pk}, yielding
%
\begin{equation} \label{eq:akhat_in_terms_of_pkhat}
    \hat{a}_k = \frac{p(k)+d_k}{\sqrt{\text{tr}(E_k)}} - \frac{\sqrt{\text{tr}(E_k)}}{d} = a_k + e_k,
\end{equation}
where $e_k = d_k/\sqrt{\text{tr}(E_k)}$ for $1 \leq k \leq M$. Finally, an estimate $(\hat{\rho}-I/d)$ of $(\rho-I/d)$ is constructed by replacing the $\{a_k\}$ with the $\{\hat{a}_k\}$ in Equation \ref{eq:rho_decomposition_no_error},
\begin{equation}
    \hat{\rho}-\frac{I}{d} = \sum_{k=1}^M \hat{a}_k \, \tilde{Q}_k = \rho-\frac{I}{d}+\rho_e,
\end{equation}
where we have defined $\rho_e = \hat{\rho}-\rho = \sum_k e_k \tilde{Q}_k$. The objective is to find the synthesis frame $\{\tilde{Q}_k\}$ that minimizes the expected squared norm of $||\rho_e||$, i.e., to minimize $\mathcal{E}$ where
\begin{equation} \label{eq:mathcal_E_quantum}
    \mathcal{E} = \mathbbm{E}\left[||\hat{\rho}-I/d||^2\right].
\end{equation}

Unlike in Section \ref{sssec:application_to_entfs}, setting $\{\tilde{Q}_k\}$ equal to the canonical dual of $\{Q_k\}$ is not necessarily optimal in terms of minimizing $\mathcal{E}$ because the error values $\{e_k\}$ are not pairwise uncorrelated. In fact, it can be shown that
\begin{subequations} \label{eq:actual_ek_properties}
\begin{alignat}{1}
    \mathbbm{E}[e_k] &= 0, \quad 1 \leq k \leq M \\[5pt]
    \mathbbm{E}[e_je_k] &= \begin{cases} \frac{p(k)\,(1-p(k))}{L\,\text{tr}(E_k)} = \Delta_k^2 \text{ if } j = k \\[7pt] \frac{-p(j)\,p(k)}{\sqrt{\text{tr}(E_j)\text{tr}(E_k)}} \left(\frac{L-1}{L}\right) \text{ if } j \neq k \end{cases}, \quad 1 \leq j, \, k \leq M.
\end{alignat}
\end{subequations}
Details are given in Appendix \ref{sec:distribution_of_relative_frequencies}. The optimal synthesis frame $\{\tilde{Q}_k\}$ could be found by first whitening the $\{e_k\}$ and then computing the canonical dual of the effective analysis frame. However, as we will demonstrate in Example \ref{ex:oversampling_for_estimation}, the conclusion reached in Section \ref{ssec:robustness_of_frame_representations} under the assumption that the $\{e_k\}$ are uncorrelated is still useful in the sense that it is supported by our simulations. This essentially implies that the correlations present in the simulations are small enough that they can be disregarded for the purpose of high-level predictions and modeling. According to Equation \ref{eq:snr_oversampling_diff_variances}, the value $\mathcal{E}_\text{can}$ of $\mathcal{E}$ obtained by setting $\{\tilde{Q}_k = Q_k/C\}$ is the canonical dual frame of $\{Q_k\}$ is
\begin{equation} \label{eq:snr_oversampling_quantum}
    \mathcal{E}_\text{can} = \frac{N^2}{M^2 a^2} \sum_{k=1}^M \Delta_k^2 = \frac{N^2}{L\,M^2\,a^2} \sum_{k=1}^M \frac{p(k)\,(1-p(k))}{\text{tr}(E_k)}.
\end{equation}
%
%


\begin{example} \label{ex:oversampling_for_estimation} \normalfont
We demonstrate the utility of Equation \ref{eq:snr_oversampling_quantum} for estimating the state of a qubit. In the following simulations, we used tight IC POVMs $\{E_k\}$ corresponding to Platonic solids with $M = 4, 6, 8, 12$ vertices. For simplicity we additionally imposed the constraint that $\text{tr}(E_k) = 2/M$ for all $1 \leq k \leq M$. The value of $a$ is chosen so that all of the $\{E_k\}$ are positive semidefinite and satisfies $a^2 \propto M$. Equation \ref{eq:snr_oversampling_quantum} suggests that with all else fixed, we would expect $\mathcal{E}_\text{can}$ to scale as $1/(LM^2)$. In other words, there is a tradeoff between the number $M$ of POVM elements and the collection size $L$, which influences the magnitude of the error. Indeed, this is what we observe.

For the purposes of illustration, we chose the density operator $\rho$ to be $\rho = \ket{\psi}\bra{\psi}$ with $\ket{\psi} = \cos(\theta/2)\ket{0} + \sin(\theta/2)\ket{1}$ and $\theta = 2\pi/3$. The collection sizes used were $L = 5, 10, 50$. For a given value of $L$ and a given POVM $\{E_k\}$ with $M$ elements, we performed 500 independent trials of the following procedure: First we drew $L$ independent samples from the true probability distribution $\{p_k\}$ in order to simulate an experiment in which $L$ quantum measurements, each with POVM $\{E_k\}$, were performed on $L$ identically prepared particles in the state $\rho$. This resulted in a collection of relative frequencies $\{\hat{p}_k\}$ from which we constructed an estimated value of $(\hat{\rho}-I/d)$ and thus an error operator $\rho_e = \hat{\rho}-\rho$. Finally, the value of $||\rho_e||^2$ was computed at the end of each trial. The compilation of these values after all trials were complete were used to compute estimates of the expected value $\mathcal{E}_\text{can} = \mathbbm{E}[||\rho_e||^2]$ and the variance $\text{var}(||\rho_e||^2)$.

The mean values $\mathcal{E}_\text{can} = \mathbbm{E}[||\rho_e||^2]$ and standard deviations $\text{var}(||\rho_e||^2)^{-1/2}$ over all trials and for all combinations of $M$ and $L$ are presented in Table \ref{tab:mean_and_std} and shown in Figure \ref{fig:rho_minus_rho_hat}. The results clearly indicate that there is a tradeoff between $L$ and $M$: For fixed values of $M$, increasing the value of $L$ reduces both the mean value and standard deviation of $||\rho_e||^2$. And, for fixed values of $L$, increasing the value of $M$ also reduces the mean value and standard deviation of $||\rho_e||^2$. However, the tradeoff is not entirely symmetric. For fixed values of $M$, doubling the value of $L$ roughly halves both the mean value and standard deviation of $||\rho_e||^2$. In other words, both quantities are roughly inversely proportional to $L$. But for fixed values of $L$, doubling the value of $M$ causes the mean value and standard deviation of $||\rho_e||^2$ to become reduced nearly by a factor of 4, suggesting that the two quantities are roughly inversely proportional to $M^2$.

\begin{table}
\centering
\caption{Mean values and standard deviations of $||\rho_{se}||^2$, rounded to the nearest tenth, over all trials and for all combinations of $M$ and $L$.}
\begin{tabular}{@{} *{7}{c} @{}}
\headercell{Number of POVM \\ Elements ($M$)} & \multicolumn{3}{c@{}}{Ensemble Size ($L$)}\\
\cmidrule(l){2-4}
& 5 &  10 & 50   \\ 
\midrule
  4  & 23.8 $\pm$ 17.0 &  11.2 $\pm$ 8.2 &  2.3 $\pm$ 1.9 \\
  6  & 9.5 $\pm$ 6.7 & 5.4 $\pm$ 4.2 &  1.0 $\pm$ 0.8 \\
  8  & 5.3 $\pm$ 3.9 &  2.6 $\pm$ 2.0 &  0.5 $\pm$ 0.4 \\
  12 & 2.4 $\pm$ 1.8 &  1.2 $\pm$ 0.9 &  0.2 $\pm$ 0.2 \\
\end{tabular}
\label{tab:mean_and_std}
\end{table}

\begin{figure}
    \centering
    \includegraphics[width=4in]{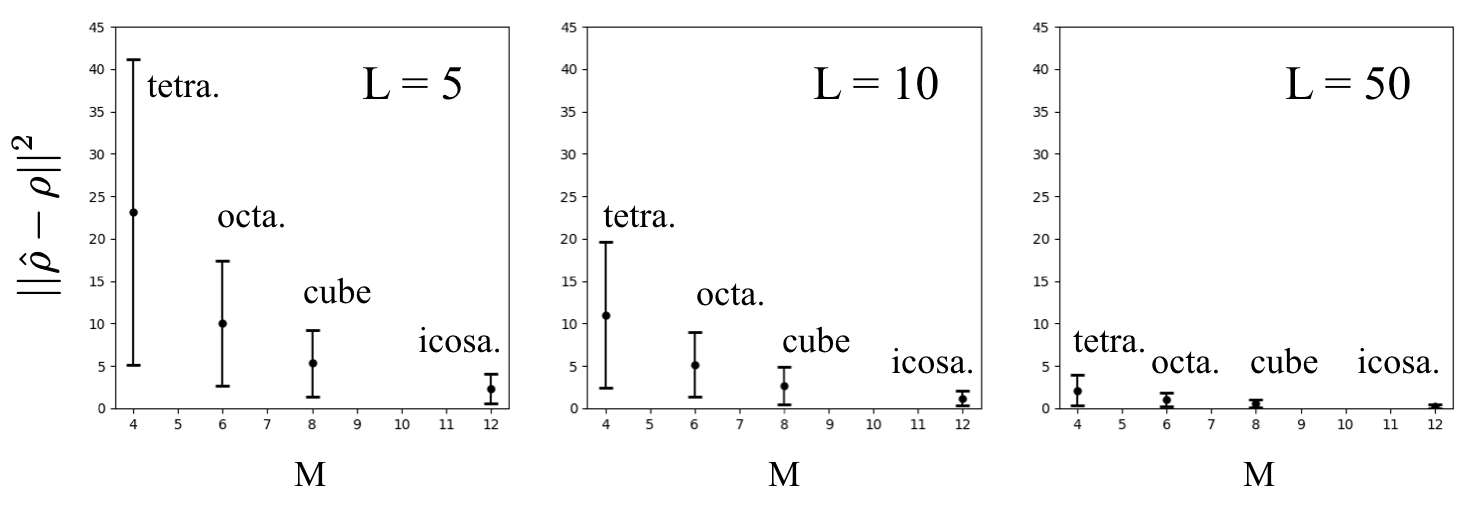}
    \caption{Mean (circular marker) and standard deviation (upper and lower error bars) of $||\hat{\rho}-\rho||^2$ over 500 independent trials for ensemble sizes $L = 5, 10, 50$ and tight IC POVMs constructed from Platonic solids with $M = 4$ (tetrahedron), $M = 6$ (octahedron), $M = 8$ (cube), and $M = 12$ (icosahedron) vertices.}
    \label{fig:rho_minus_rho_hat}
\end{figure}
\end{example}

\subsection{Quantum Binary Detection} \label{ssec:application_to_detection}


It is reasonable to assume given the analysis in Section \ref{ssec:application_to_estimation} that an analogous tradeoff between the collection size $L$ and number of POVM elements $M$ would be present in the context of quantum binary state detection. In Example \ref{ex:oversampling_for_detection} we present evidence through simulation that this is indeed the case. Namely, for fixed values of $L$ increasing the value of $M$ leads to a smaller probability of error, and vice versa. We emphasize that the version of the quantum binary state detection problem used here has many generalizations and extensions.

The formulation of the problem that we use is as follows. We start with $L$ identically prepared qubits whose state is described by one of two density operators,
\begin{equation}
    \rho = \begin{cases} \rho_0 = \ket{\psi_0}\bra{\psi_0} \text{ if } H = H_0 \\[5pt] \rho_1 = \ket{\psi_1}\bra{\psi_1} \text{ if } H = H_1 \end{cases}
\end{equation}
with prior probabilities $P(H = H_0) = q_0$ and $P(H = H_1) = q_1$. We discriminate between the two possibilities by first choosing a fixed POVM $\{E_k\}$ and performing a quantum measurement whose associated POVM is $\{E_k\}$ on each of the $L$ particles. This results in a relative frequency vector $\hat{p} = [\hat{p}_1, \dots, \hat{p}_M]^T$. We then perform a likelihood ratio test (LRT) on $\hat{p}$ with threshold $\eta = q_0/q_1$ in order to make a final decision. It is well-known that if the only information on which to base a decision is the relative frequency vector, this decision strategy minimizes the probability of error.

\begin{example} \label{ex:oversampling_for_detection} \normalfont
We arbitrarily set $\rho_0 = \ket{0}\bra{0}$ and $\rho_1 = \ket{\psi}\bra{\psi}$ where $\ket{\psi} = \cos(\theta/2)\ket{0} + e^{j\phi}\sin(\theta/2)\ket{1}$ with $\theta = 2\pi/3$ and $\phi = \pi/3$. For a set of POVMs corresponding to Platonic solids with $M = 4, 6$ vertices and for collection sizes of $L = 5, 10, 20$, we performed LRTs with thresholds ranging from 0 to $\infty$ on the relative frequency vectors and plotted the corresponding values of the probability of false alarm ($P_f$) and the probability of detection ($P_d$). Each threshold corresponds to the minimum probability of error decision strategy for some combination of prior probabilities. The results are shown in Figure \ref{fig:platonic_solids_lrt_qdocs}.In the terminology of [Allerton, FnT], the curves are referred to as LRT QDOCs. The three plots reflect the anticipated tradeoff between $M$ and $L$. For a fixed value of $L$, increasing the value of $M$ (informally, the level of overcompleteness of the POVM) leads to better detection as reflected by the superior QDOC. On the other hand, for a fixed value of $M$ increasing the value of $L$ also leads to better detection.

\begin{figure}
    \centering
    \includegraphics[width=4in]{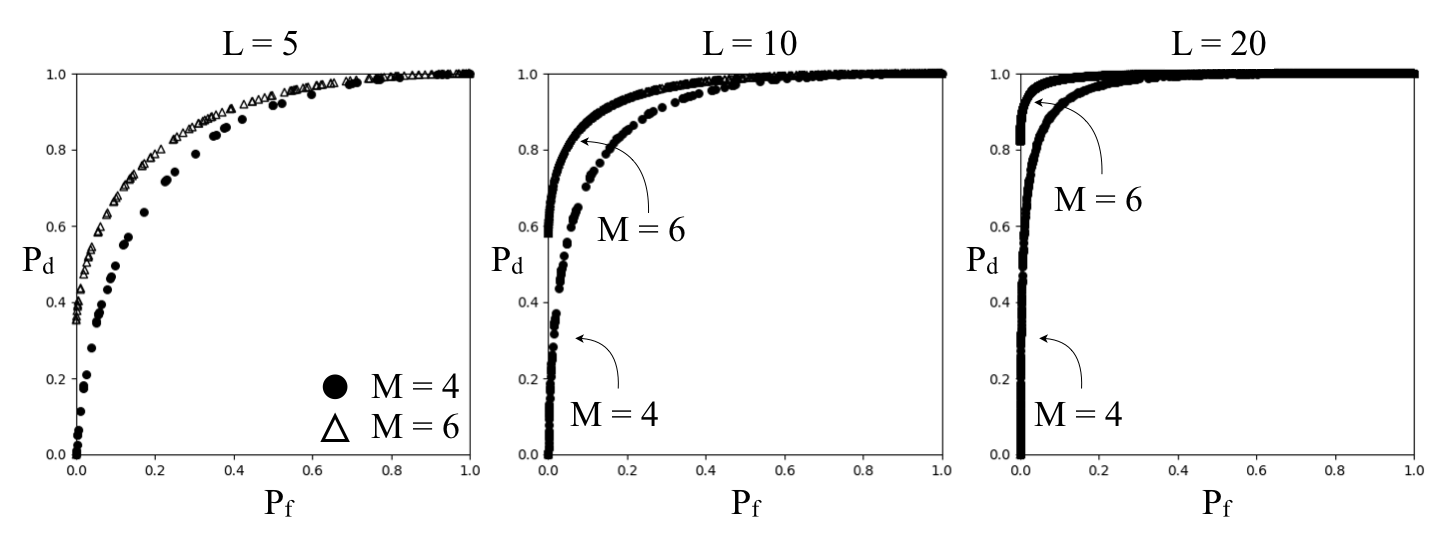}
    \caption{LRT QDOCs for ensemble sizes $L = 5, 10, 15$ and tight IC POVMs constructed from Platonic solids with $M = 4$ (tetrahedron) and $M = 6$ (octahedron) vertices.}
    \label{fig:platonic_solids_lrt_qdocs}
\end{figure}
\end{example}

\begin{example} \label{ex:overcompleteness_stability_of_orientation} \normalfont
Finite sample size estimations of the probabilities are one of many sources of error that may affect detection performance. Another is preparation noise affecting the value of $\rho$, effectively causing its true Bloch vector to be misaligned, or rotated some amount within the Bloch sphere. This may equivalently be thought of as misalignment of the Bloch vectors of the POVM elements. In the consideration of POVMs constructed from Platonic solids, POVMs with more elements are more robust to this type of error since, roughly speaking, the relative orientation of the Platonic solid to the Bloch vectors of $\rho_0$ and $\rho_1$ is more stable when the solid is a closer approximation of a sphere. However, increasing the number $M$ of POVM elements is not without cost. From our observations the minimum probability of error that is achievable by any orientation of a given Platonic solid increases with $M$. In other words, we have observed a tradeoff between the best performance that can be achieved by a given Platonic solid, and the robustness it possesses in relation to that performance. This is shown in Figure \ref{fig:overcompleteness_stability_of_orientation} for three Platonic solids with $M = 4, 6, 8$ vertices as well as a degenerate solid consisting of two antipodal points on the Bloch sphere. The degenerate solid corresponds to a standard quantum measurement. To generate each subplot, we used the same $\rho_0$ and $\rho_1$ as in Example \ref{ex:oversampling_for_detection} as well as priors of $q_0 = q_1 = 1/2$. For each orientation of a given Platonic solid on the Bloch sphere, we computed the probability of error corresponding to an LRT with threshold $\eta = q_0/q_1 = 1$. Each subplot shows the variation in this probability of error for a range of orientations. As expected, higher values of $M$ display a lower sensitivity to misalignment on the Bloch sphere but also higher global minima in the probability of error over all alignments.

\begin{figure}
    \centering
    \includegraphics[width=6in]{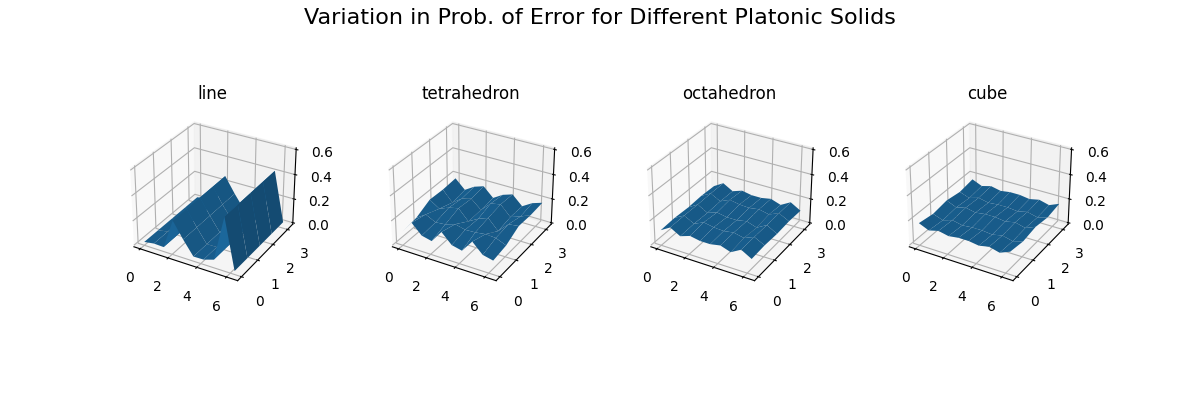}
    \caption{Minimum probability of error achievable for each of a range of orientations of different Platonic solids.}
    \label{fig:overcompleteness_stability_of_orientation}
\end{figure}
\end{example}

\appendix

\section{Optimality of the Canonical Dual} \label{sec:optimality_of_canonical_dual}

The derivation given below does not assume that the analysis frame is an ENTF. The problem described in Section \ref{sssec:problem_description} can be formulated as
%
%
\begin{mini!}|l|[2] 
    {L} 
    {|| L( \, e \, ) ||^2 \label{eq:cost_function}} 
    {\label{eq:optimization}} 
    {} 
    \addConstraint{L\,(\,A\,(\,v\,)\,)}{= \ket{v} \text{for all } \ket{v} \in \mathcal{V} \label{eq:left_inverse}} 
\end{mini!}
where the minimization is performed over all linear operators $L$ from $\mathcal{W}$ to $\mathcal{V}$. The constraint, which in effect specifies that $L$ must be the synthesis operator of a frame that is dual to the analysis frame, amounts to the requirement that $L$ is a left-inverse of $A$. A left-inverse is guaranteed to exist because as stated in Section \ref{ssec:analysis_and_synthesis_operators}, $A$ has rank $N$.

Let $L$ be an arbitrary left-inverse of $A$ and assume that $\{\vec{w}_k, \, 1 \leq k \leq M\}$ is an orthonormal basis (ONB) for $\mathcal{W}$. Further assume that the $\{\vec{w}_k\}$ can be partitioned into an ONB $\{\vec{w}_k, \, 1 \leq k \leq N\}$ for $\text{range}(A)$ and an ONB $\{\vec{w}_k, \, N+1 \leq k \leq M\}$ for $\text{range}(A)^\perp$. To fully specify the operator $L$, it is both necessary and sufficient to specify its action on each of the $\{\vec{w}_k\}$. Briefly, its action on $\text{range}(A)$ must be chosen to satisfy the constraint \ref{eq:left_inverse} while its action on $\text{range}(A)^\perp$ can be chosen to minimize $||L(e)||^2$.

We first consider its action on $\text{range}(A)$. For each $\{\vec{w}_k, \, 1 \leq k \leq N\}$, there is a unique vector $\ket{v_k} \in \mathcal{V}$ satisfying $A(v_k) = \vec{w}_k$. The requirement \ref{eq:left_inverse} implies that $L(w_k) = \ket{v_k}$ for all $1 \leq k \leq N$. The action of $L$ on $\text{range}(A)^\perp$ can now be chosen to minimize $||L(e)||^2$. Note that any error vector $\vec{e} \in \mathcal{W}$ can be written uniquely as
\begin{subequations}
\begin{alignat}{1}
    \vec{e} &= \vec{e}_1 + \vec{e}_2 = \sum_{k=1}^N c_k \, \vec{w}_k + \sum_{k=N+1}^M c_k \, \vec{w}_k \\[5pt]
    \vec{e}_1 &\in \text{range}(A) \\[5pt]
    \vec{e}_2 &\in \text{range}(A)^\perp,
\end{alignat}
\end{subequations}
where $\{c_k\}$ are the coefficients of $\vec{e}$ in the $\{\vec{w}_k\}$ basis. It is straightforward to show that since the $\{c_k\}$ are related to the $\{e_k\}$ by an orthogonal transformation in $\mathcal{W}$, they also have zero mean, variance $\sigma^2$, and are pairwise uncorrelated. The expected value of $||L(e)||^2$ is
\begin{subequations}
\begin{alignat}{1}
    \mathbbm{E}\left[||L(e)||^2\right] = \mathbbm{E}\left[||L(e_1)+L(e_2)||^2\right].
\end{alignat}
\end{subequations}
As we will show below, the expected value is minimized when $L(e_2)$ is set to zero for all values of $\vec{e}_2$. The vector $L(e)$ is equal to
\begin{subequations}
\begin{alignat}{1}
    L(e) &= \sum_{k=1}^N c_k\,L(\vec{w}_k) + \sum_{k=N+1}^M c_k\,L(\vec{w}_k) \\[7pt]
    &= \sum_{k=1}^N c_k\,\ket{v_k} + \sum_{k=N+1}^M c_k\,L(\vec{w}_k).
\end{alignat}
\end{subequations}
Its squared norm is equal to $\braket{L(e)|L(e)}$, and since the $\{c_k\}$ are pairwise uncorrelated all cross terms are equal to zero. Thus,
\begin{subequations}
\begin{alignat}{1}
    \mathbbm{E}\left[||L(e)||^2\right] &= \mathbbm{E}\left[ \sum_{k=1}^N c_k^2\,||v_k||^2 + \sum_{k=N+1}^M c_k^2\,||L(\vec{w}_k)||^2 \right] \\[7pt]
    &= \sum_{k=1}^N \mathbbm{E}[c_k^2]\,||v_k||^2 + \sum_{k=N+1}^M \mathbbm{E}[c_k^2]\,||L(\vec{w}_k)||^2 \\[7pt]
    &= \sigma^2 \sum_{k=1}^N ||v_k||^2 + \sigma^2 \sum_{k=N+1}^M ||L(\vec{w}_k)||^2.
\end{alignat}
\end{subequations}
Since the value of the first sum is fixed and since all terms in both sums must be non-negative, the minimal value is obtained when the second sum is equal to zero, which happens when $L(\vec{w}_k) = 0$ for all $N+1 \leq k \leq M$. Thus, the optimal left-inverse $L^*$ inverts $A$ over its range and acts as the zero operator on $\text{range}(A)^\perp$. It is well-known (see, for example, Chapter 1 of \cite{casazza2012finite}) that the unique left-inverse with these properties is the Moore-Penrose pseudoinverse of $A$. Explicitly, the pseudoinverse is equal to
\begin{equation}
    L^* = (A^\dagger A)^{-1} A^\dagger,
\end{equation}
and this corresponds exactly to the synthesis operator of the canonical dual frame \cite{casazza2012finite}.

\section{Distribution of Relative Frequencies} \label{sec:distribution_of_relative_frequencies}

While the following derivation is motivated by the problem the quantum state estimation problem considered in Section \ref{ssec:application_to_estimation}, the concepts and conclusions are not reliant on the postulates of quantum mechanics. To emphasize this point we state the results without any reference to density operators or quantum measurement. Let $X$ be a discrete random variable that takes values in the set $\{1, \dots, M\}$ with probability mass function (PMF) $\{p(1), \dots, p(M)\}$, i.e.,
\begin{equation}
    X = k \text{ with probability } p(k), \, 1 \leq k \leq M.
\end{equation}
Assume that $\{x_i, 1 \leq i \leq L\}$ is a set of $L$ independent realizations of $X$ and consider the set of relative frequencies $\{\hat{p}(k) = \ell_k/L\}$, where $\ell_k$ is the number of realizations $\{x_i\}$ that are equal to $k$. Defining $d_k = p(k) - \hat{p}(k)$ for $1 \leq k \leq M$, we wish to evaluate the quantities $\mathbbm{E}[d_k]$ and $\mathbbm{E}[d_j d_k]$ for arbitrary values of $1 \leq j, \, k \leq M$.

Let $k$ be a fixed but arbitrary integer between 1 and $M$. To compute the expected value $\mathbbm{E}[d_k]$, note that the value of $\ell_k$ is binomially distributed with parameters $p(k)$ and $L$. Its expected value is $\mathbbm{E}[\ell_k] = L \, p(k)$ and its variance is $\text{var}(\ell_k) = L \, p(k) \, (1-p(k))$. Using linearity of expectation we find that, unsurprisingly, the expected value of $d_k$ is equal to zero,
\begin{equation}
    \mathbbm{E}[d_k] = \mathbbm{E}\left[p(k) - \frac{\ell_k}{L}\right] = p(k) - \frac{L \, p(k)}{L} = 0.
\end{equation}
The variance of $d_k$ is
\begin{equation}
    \text{var}(d_k) = \text{var}\left(p(k) - \frac{\ell_k}{L}\right) = \frac{\text{var}(\ell_k)}{L^2} = \frac{p(k)\,(1-p(k))}{L}.
\end{equation}
Furthermore, since $\mathbbm{E}[d_k] = 0$ we have $\text{var}(d_k) = \mathbbm{E}[d_k^2]$.

Now let $j$ and $k$ be fixed but arbitrary integers between 1 and $M$ with $j \neq k$. To compute the value of $\mathbbm{E}[\Delta_j\Delta_k]$, note that the joint distribution of $\{\ell_1, \dots, \ell_M\}$ is given by a multinomial distribution with parameters $L$ and $\{p_1, \dots, p_M\}$.  Specifically, given a set of non-negative integers $\{\ell_1, \dots, \ell_M\}$ that sum to $L$, the probability that $\ell_k$ of the $\{X_i\}$ are equal to $k$ for $1 \leq k \leq M$ is
\begin{equation} \label{eq:pmf_of_ell}
f(\ell_1, \dots, \ell_M) = C_L(\ell_1, \dots, \ell_M) \, \prod_{k=1}^M p(k)^{\ell_k}.
\end{equation}
In Equation \ref{eq:pmf_of_ell}, $f(\cdot)$ is used to denote the joint distribution of the $\{\ell_k\}$ and the constant $C_L(\ell_1, \dots, \ell_M)$ is a combinatorial factor that accounts for the fact that the ordering of the realizations does not affect the values of the $\{\ell_k\}$. We have
\begin{equation}
    C_L(\ell_1, \dots, \ell_M) = {L \choose \ell_1} \, {L-\ell_1 \choose \ell_2} \, \dots \, {L-\ell_1 - \dots - \ell_{M-2} \choose \ell_{M-1}} = \frac{L!}{\ell_1! \dots \ell_M!}
\end{equation}
It is well-known and can be shown using the properties of the multinomial distribution that
\begin{equation}
    \mathbbm{E}[\ell_j \ell_k] = -L \, p(j) \, p(k).
\end{equation}
Using linearity of expectation and the fact that $\mathbbm{E}[\ell_j] = L\,p(j)$ and $\mathbbm{E}[\ell_k] = L\,p(k)$, we find that the value of $\mathbbm{E}[d_j d_k]$ is
\begin{subequations}
\begin{alignat}{1}
    \mathbbm{E}[d_j d_k] &= \mathbbm{E}\left[\left(p(j)-\frac{\ell_j}{L}\right)\left(p(k)-\frac{\ell_k}{L}\right)\right] = - p(j) \, p(k) + \frac{\mathbbm{E}[\ell_j \ell_k]}{L^2} \\[7pt]
    &= -p(j) \, p(k) - \frac{p(j)\,p(k)}{L} = - p(j) \, p(k) \, \left(\frac{L-1}{L}\right).
\end{alignat}
\end{subequations}

\bibliographystyle{IEEEbib}
\bibliography{refs}

\end{document}